\documentclass[twocolumn,aps,tightenlines,floatfix,showpacs]{revtex4}
\usepackage[dvips]{graphicx}
\usepackage{hhline}

\newcommand{\be}{\begin{equation}}
\newcommand{\ee}{\end{equation}}

\begin{document}
\title{High-Performance Algorithm for Calculating Non-Spurious Spin- and
Parity-Dependent Nuclear Level Densities}

\author{R.A. Sen'kov$^1$, M. Horoi$^1$, and V.G. Zelevinsky$^2$}

\affiliation{$^1$Department of Physics, Central Michigan University, Mount Pleasant, Michigan 48859, USA%}
%\affiliation{
\vspace*{0.3cm}\\
$^2$Department of Physics and Astronomy and National Superconducting Cyclotron Laboratory, Michigan State University, East Lansing, MI 48824-1321,USA }


\begin{abstract}
A new high-performance algorithm for calculating the spin- and parity-dependent shell model nuclear level
densities using methods of statistical spectroscopy in the proton-neutron formalism was recently proposed. 
When used in valence spaces that cover more than one major harmonic oscillator shell, this algorithm mixes 
the genuine intrinsic states with spurious center-of-mass excitations. In this paper we present an advanced 
algorithm, based on the recently proposed statistical moments method, that eliminates the spurious states. 
Results for unnatural parity states of several $sd$-shell nuclei are presented and compared with those of 
exact shell model calculations and experimental data.

\end{abstract}

\maketitle

\section{Introduction}

In this article we make a next step towards a reliable practical
algorithm for calculating the level density in a finite many-body system
with strong interaction between the constituents. Our primary object of
applications is the atomic nucleus but the same techniques can be applied
to other mesoscopic systems, such as atoms in traps \cite{heiselberg02}
and quantum dots \cite{reimann02}. Being a key element in the description
of nuclear reactions and quantum transport in general, the many-body
level densities are also interesting from the fundamental viewpoint since,
at not very high excitation energy, the growth of the level density
reflects the interplay of interactions inside the system. The energy
dependence of the level density may indicate the phase transformations
smoothed out in finite systems: pairing quenching in nuclei \cite{big,HZppt}
and small metallic particles \cite{leavens81}, or magnetic effects in
small quantum dots \cite{gonzalez02}. The correct reproduction of the level
density serves as the first step to recognizing the regular or chaotic
nature of spectral statistics in nuclei \cite{brody81,big,aberg09}, complex
atoms \cite{grib} and quantum dots \cite{ulloa97}.

The information on spin- and parity-dependent nuclear level densities (SPNLD)
represents a critical ingredient for the theory of nuclear reactions, including
those of astrophysical interest. For example, the routinely used Hauser-Feschbach
approach \cite{HF} requires the knowledge of nuclear level densities for certain
quantum numbers $J^{\pi}$ of spin and parity in the window of excitation
energy around the particle threshold \cite{rauscher97,moller09}. The nuclear
technology requires the level density in the region of compound resonances
close to the neutron separation energy. Therefore, a lot of effort has been
invested in finding the accurate SPNLD, starting with the classical Fermi-gas
approximations \cite{bethe36,cameron,ericson} and progressing to the sophisticated
mean-field combinatorics \cite{gor-icomb-08,gor-fis-09,aberg09} and various shell model
approaches with residual interactions in large valence spaces
\cite{Eric97,Alhassid97,Langanke98,alhassid00,jtpden,jdenrc,alhassid07,alhassid09,SH2010}.

Earlier we developed a strategy \cite{jtpden,jdenrc,nic8den,epl10,SH2010} of
calculating the shell model SPNLD using methods of statistical spectroscopy
\cite{wongbook,kotabook}. In the basic version of this approach, all basis states
that can be built within a selected spherical single-particle valence space are
taken into account. However, when the valence space spans more than one harmonic
oscillator major shell, the shell model states include spurious excitations of
the center-of-mass (CM). Apart from the level density problem, the correct
separation of the spurious states is important for finding the physical response
of the system to external fields, for example in the case of the excitation of
isoscalar giant resonances. Over the years, shell model practitioners developed
techniques \cite{lawson} that allow one to push these unwanted states to energies
higher than those of interest for low-energy phenomena. These techniques involve
(i) separation of basis states in the so-called $N\hbar\omega$ subspaces that
can exactly factorize in the CM-excited and intrinsic states, and (ii) the addition
to the nuclear Hamiltonian of a CM-part multiplied by a properly chosen positive
constant. Unfortunately, the second ingredient leads to a multimodal distribution
of levels and it is not appropriate for the methods of statistical spectroscopy.

In a recent letter \cite{dencom}, we proposed a shell-model algorithm for removing
the spurious CM contributions from the SPNLD that works if one knows the SPNLD
for each $N\hbar\omega$ subspace. These contributions are unimodal and can be
described using methods of statistical spectroscopy, provided that one can
calculate the necessary moments of the Hamiltonian in $N\hbar\omega$ subspaces.
In the present paper we formulate and utilize a high-performance algorithm that
can calculate the configuration centroids and the widths of the Hamiltonian in
$N\hbar\omega$ subspaces. This algorithm will be used for calculating the
non-spurious level density for unnatural parity states of several $sd$-shell nuclei.
Comparison with the exact shell model results and/or experimental data will be
also presented.

\begin{figure*}
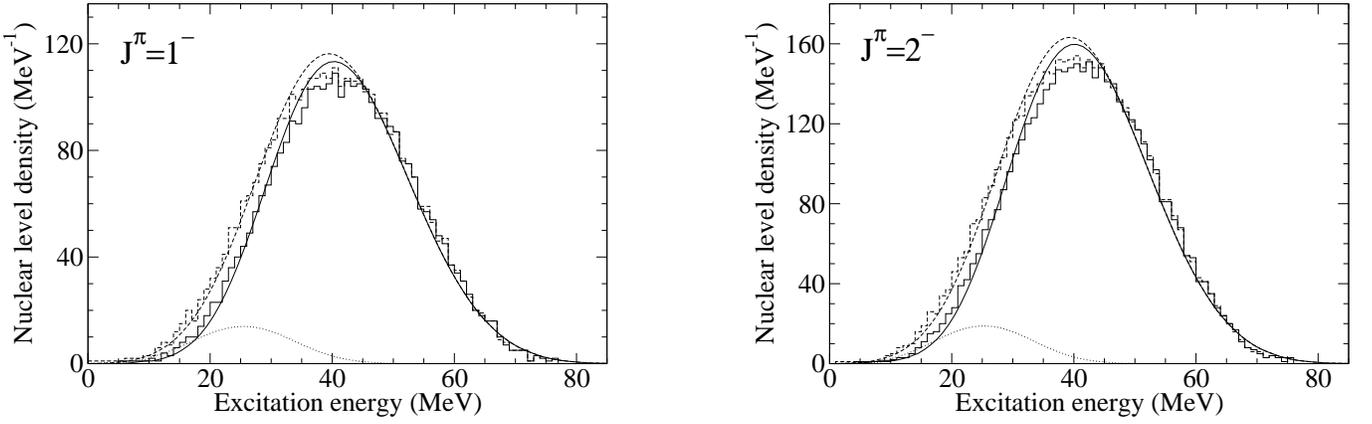

\includegraphics[width=0.45\textwidth]{pict/ne20/ne20_1.eps}
\hfill
\includegraphics[width=0.45\textwidth]{pict/ne20/ne20_2.eps}

\caption{${}^{20}$Ne, $1\hbar\omega$, negative parity. Comparison of nuclear level densities
from the exact shell model (stair lines) and from the moments method (straight lines).
Dashed lines correspond to total densities with spurious states included; solid lines
correspond to non-spurious densities without spurious states. Dotted lines present the
spurious density.}
\end{figure*}

\section{Spin- and Parity-dependent configuration moments method}

We start closely following the approach proposed in Refs. \cite{jtpden,jdenrc,SH2010}.
For reader's convenience, we repeat first the main equations of the moments
method. According to this approach, one can calculate the level density $\rho$ as
a function of excitation energy $E$ in the following way:
\be \label{rho1}\label{1}
\rho(E,\alpha) = \sum_\kappa D_{\alpha \kappa } \cdot G_{\alpha \kappa} (E),
\ee
where $\alpha = \{n, J, T_z, \pi \}$ is a set of quantum numbers, the total number
of particles, $n$, total spin, $J$, isospin projection, $T_z$, and parity, $\pi$;
for the level density, in contrast to the state density, the spin degeneracy $2J+1$
is excluded. The sum over configurations $\kappa$ in Eq. (\ref{rho1}) spans all
possible (for the certain values of $n,T_z$, and $\pi$) ways of distributing $n$
particles over $q$ spherical single-particle orbitals. Each configuration $\kappa$
is presented by a set of occupation numbers $\kappa=\{n_1, n_2,... \; , n_q \}, $
where $n_j$ is the number of particles occupying the spherical single-particle level
$j$, $\sum_{j}n_{j}=n$.

The energy dependence of the density $\rho$ is expressed by finite-range Gaussian
functions, $G_{\alpha \kappa}$, defined as in Ref. \cite{jtpden}:
\begin{eqnarray}\label{frg}\label{2}
G_{\alpha \kappa}(E)=G(E+E_{{\rm g.s.}}-E_{\alpha \kappa},\sigma_{\alpha \kappa}),\\
G(x,\sigma)= C \cdot \left\{
\begin{array}{ll}\label{3}
\mbox{exp}\left( -x^2/2\sigma^2 \right) &, \; \; |x| \leq \eta \cdot \sigma \\
0 &, \; \; |x|> \eta \cdot \sigma \\
\end{array} \right. ,
\end{eqnarray}
where the parameters $E_{\alpha \kappa}$ and $\sigma_{\alpha \kappa}$ will be
defined later, $E_{{\rm g.s.}}$ is the ground state energy, $\eta$ is the cut-off
parameter, and $C$ is the normalization factor corresponding to the condition 
$\int_{-\infty}^{+\infty}G(x,\sigma)dx=1$. Although $\eta$ can be treated as
a free parameter, we know from the previous works (see for example
\cite{SH2010,epl10,nic8den}) that its optimal value is $\eta\sim 3$, which allows
us to get a finite-range distribution for the density and, practically, do not change
each Gaussian contribution. Finally, the dimension $D_{\alpha \kappa}$ in Eq. (\ref{1})
gives the correct normalization for each finite-range Gaussian being equal to
the number of many-body states with given set of quantum numbers $\alpha$ that
can be built for a given configuration $\kappa$.

The density distribution $\rho(E)$, especially its low-energy part, is very
sensitive to the choice of the ground state energy $E_{gs}$. This origin of
the energy scale is an external parameter for the moments method. To calculate
it we need to use supplementary methods, such as the shell model.

The parameters $E_{\alpha \kappa}$ and $\sigma_{\alpha \kappa}$ in Eq. (\ref{frg})
are the fixed-$J$
configuration centroids and widths. They essentially present the average
energy and the standard deviation for the set of many-body states with a given
set of quantum numbers $\alpha$ within a given configuration $\kappa$.
For a Hamiltonian containing one- and (antisymmetrized) two-body parts of interaction,
\be
\label{h}\label{4}
H = \sum_i \epsilon_i a^\dag_i a_i + \frac{1}{4} \sum_{i j k l} V_{i j k l}
a^\dag_i a^\dag_j a_l a_k,
\ee
the fixed-$J$ centroids and widths can be expressed in terms of
the traces of the first and second power of this Hamiltonian \cite{SH2010},
Tr$[H]$ and Tr$[H^2]$, for each configuration $\kappa$:
\begin{eqnarray}\label{5}
E_{\alpha \kappa} = \langle H \rangle_{\alpha \kappa}, \\
\sigma_{\alpha \kappa} = \sqrt{\langle H^2\rangle_{\alpha \kappa}
- \langle H \rangle^2_{\alpha \kappa} },\label{6}
\end{eqnarray}
where
\begin{eqnarray}
\label{tr1}\label{7}
\langle H \rangle_{\alpha \kappa} =
\mbox{Tr}^{(\alpha \kappa)}[H]/D_{\alpha \kappa},\\
\label{tr2}\label{8}
\langle H^2 \rangle_{\alpha \kappa} =
\mbox{Tr}^{(\alpha \kappa)}[H^2]/D_{\alpha \kappa}.
\end{eqnarray}
Every trace, such as $\mbox{Tr}^{(\alpha \kappa)}[\cdots]$, contains the sum of all
diagonal matrix elements, $\sum\langle\nu, J | \cdots | \nu, J \rangle$, over all
many-body states $\left|\nu, J \right>$ within given configuration $\kappa$ and with
certain set of quantum numbers $\alpha$.

As in our previous work \cite{SH2010} we derive these traces for the fixed total
spin projection $J_z$, rather than for fixed $J$. Technically, it is much more easier
to construct many-body states with a given total spin projection. The $J$-traces
can be easily expressed through the $J_z$-traces using the standard relation
\be\label{9}
\label{mztoJ}
\mbox{Tr}^{(J)} [\cdots]
= \mbox{Tr}^{(J_z)} [\cdots]_{{}_{J_z=J}} -
\mbox{Tr}^{(J_z)} [\cdots]_{{}_{J_z=J+1}}.
\ee
In Eq. (\ref{mztoJ}) we omitted all quantum numbers, except for the projection $J_z$
and the total spin $J$.

\begin{figure*}
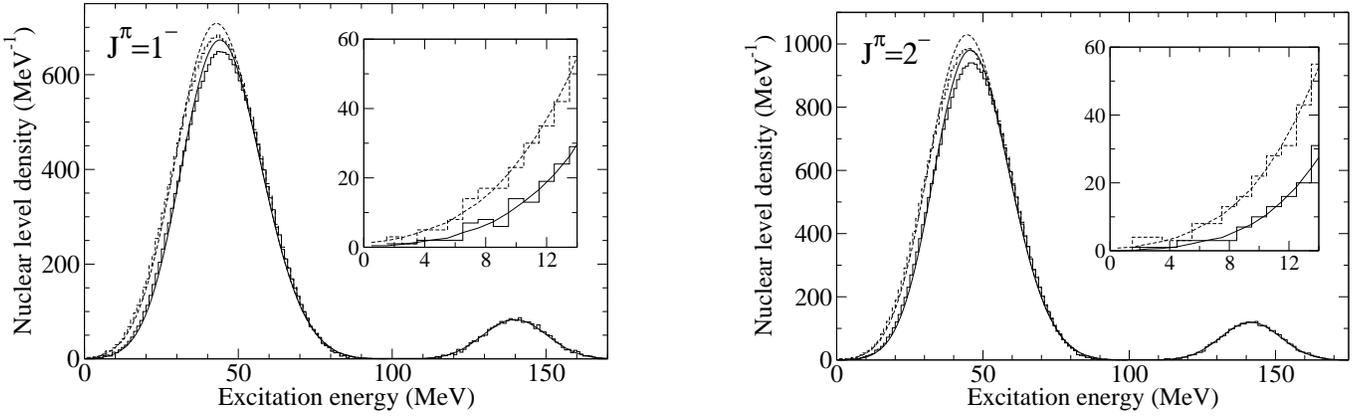

\includegraphics[width=0.45\textwidth]{pict/na22/na221hw-1n.eps}
\hfill
\includegraphics[width=0.45\textwidth]{pict/na22/na221hw-2n.eps}

\caption{${}^{22}$Na, $1\hbar\omega$, negative parity. Comparison of nuclear level densities from the exact shell model (stair lines) and from the moments method (straight lines).
Dashed lines correspond to total densities with spurious states included; solid lines
correspond to non-spurious densities without spurious states. Right side of the figures presents the spurious density. }
\end{figure*}

We will use the same label $\alpha$ to denote a set of quantum numbers that includes
either the fixed $J_z$ or the fixed $J$, keeping in mind that Eq. (\ref{mztoJ})
can always connect them. In every important case we will point out which set of
quantum numbers was used. The expressions for the traces can be found in Ref.
\cite{jaq79}, see also \cite{SH2010} and references there.

\begin{figure*}
\includegraphics[width=0.45\textwidth]{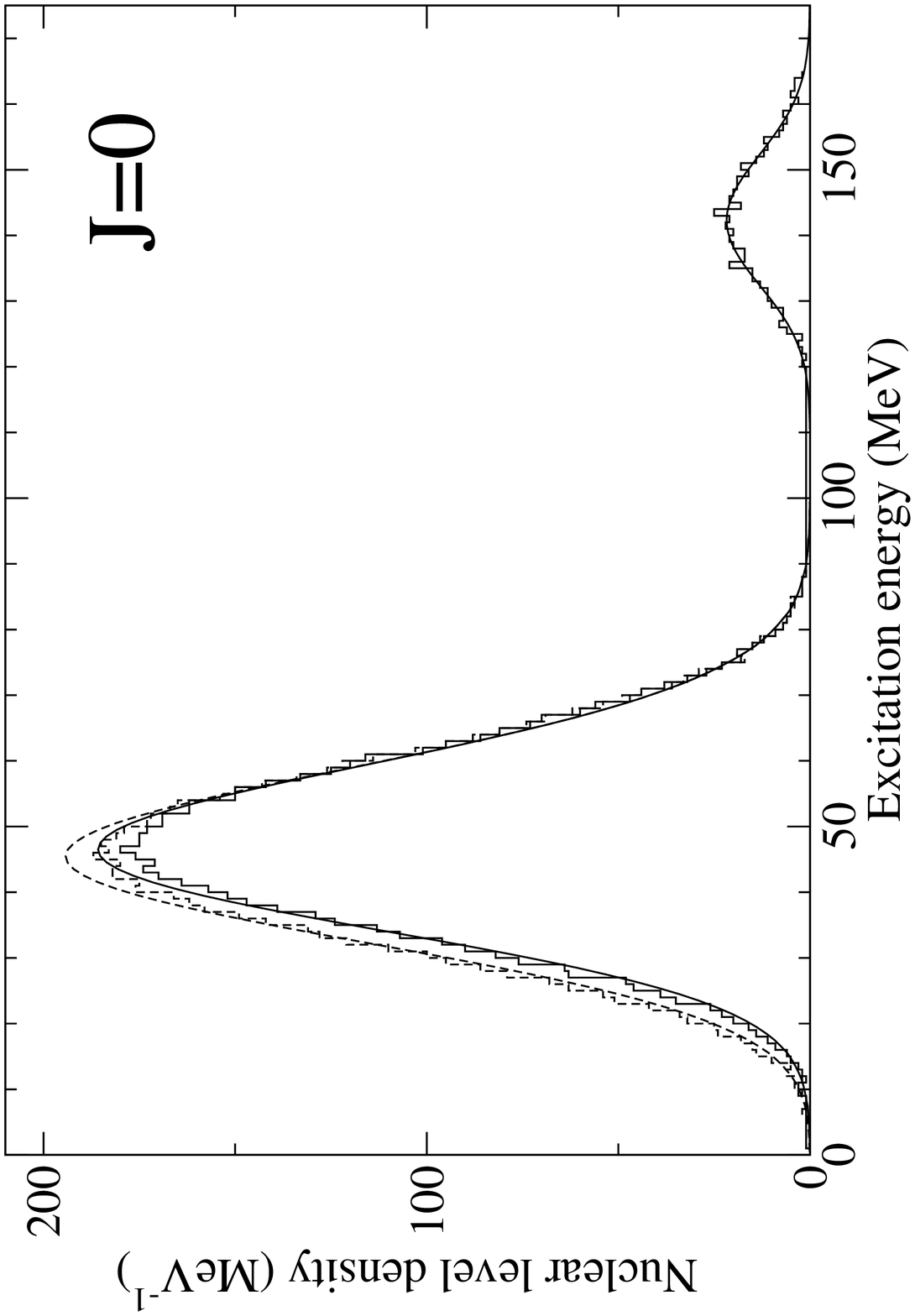}
\hfill
\includegraphics[width=0.45\textwidth]{pict/mg22/den1new.eps}

\vspace{1.0cm}

\includegraphics[width=0.45\textwidth]{pict/mg22/den2new.eps}
\hfill
\includegraphics[width=0.45\textwidth]{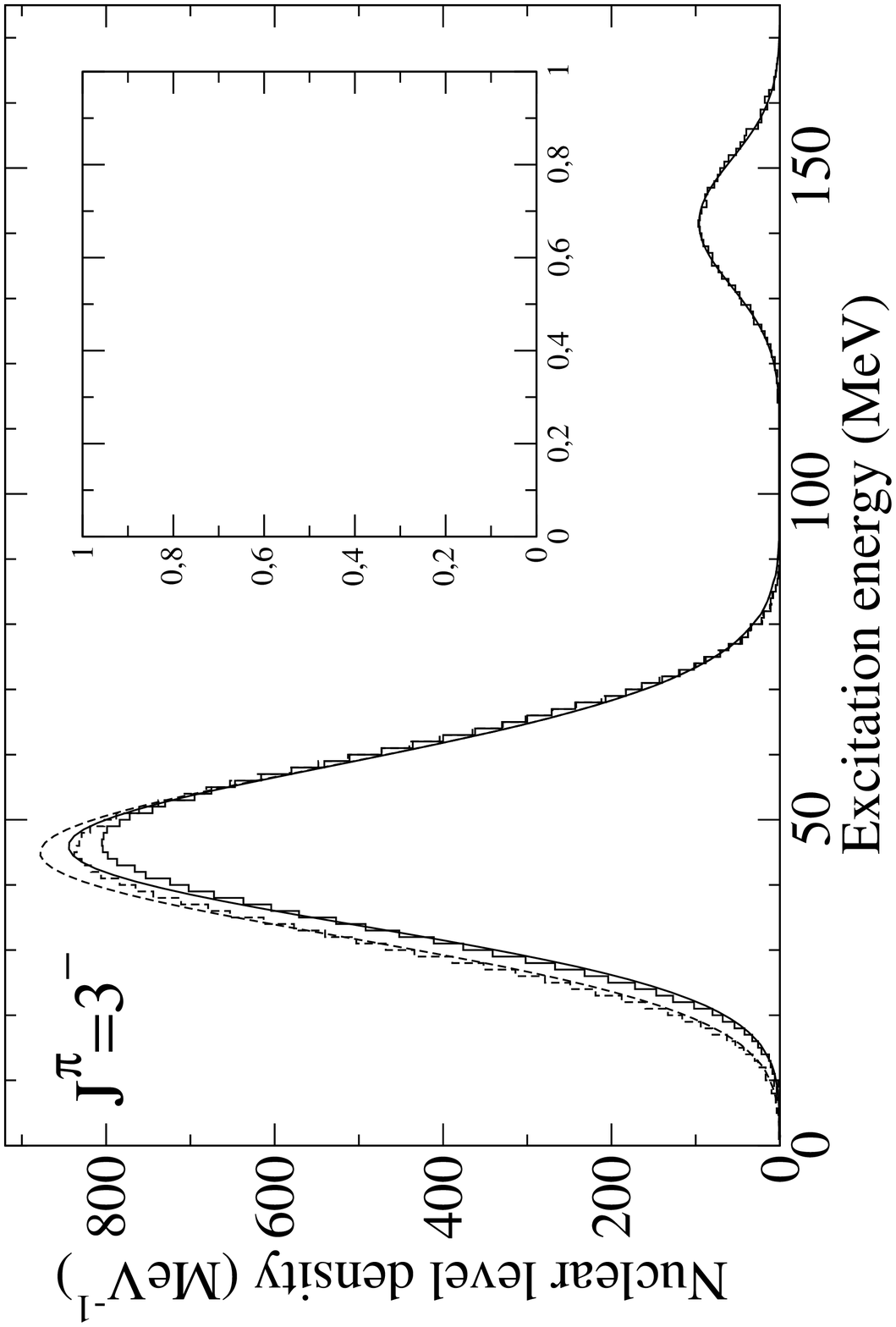}
%\hfill
%\includegraphics[width=0.32\textwidth]{pict/mg22/den4.eps}
%\hfill
%\includegraphics[width=0.32\textwidth]{pict/mg22/den5.eps}
\caption{${}^{22}$Mg, negative parity. Comparison of nuclear level densities
between exact shell model (stair lines) and moments method (straight lines).
Dashed lines correspond to total densities with spurious states. Solid lines
correspond to non-spurious densities without spurious states. Right side of the figures
present spurious density.}

\end{figure*}

\section{New features of the algorithm}

To remove the center-of-mass spurious states from the level density we follow
the approach suggested in \cite{dencom}. This approach assumes the knowledge
of the level density within a restricted basis, and one needs to calculate partial
densities $\rho(E,\alpha, N)$ for a given number $N$ of excitations
$\hbar\omega$, where the classification of states in terms of the harmonic
oscillator $N\hbar\omega$ excitations is applied. It has to be mentioned that
such a splitting is clearly an approximation since it refers to the harmonic
oscillator potential and the level scheme. The harmonic oscillator frequency $\omega$,
being an auxiliary parameter has no strict {\sl a-priori} definition. Nevertheless,
as we will demonstrate, the suggested method results are independent of $\hbar\omega$.

According to this method a pure (without admixture of spurious states) level
density $\rho^{(0)}(E,J,N)$ can be expressed through the total
(all states included) level density $\rho(E,J,N)$ for the same values
of arguments and through the pure densities of lower $N$, Ref. \cite{dencom}:
$$
\rho^{(0)}(E,J,N)=\rho(E,J,N)-
$$
\be\label{12}
-\sum_{K=1}^N \sum_{J_{K}=J_{{\rm min}}}^{N,{\rm step} \; 2}
\sum_{J'=|J-J_{K}|}^{J+J_{K}}\rho^{(0)}(E,J',(N-K)).
\ee
Here, for simplicity, we omitted all quantum numbers indicatting only total
spin $J$ representing the set $\alpha$. In order to make these recursive equations
complete we need a boundary condition that can be obtained from the $0\hbar\omega$ case
which is free of spurious admixtures,
\be\label{13}
\rho^{(0)}(E,J,0)=\rho(E,J,0).
\ee
For example, if we are interested in the $1\hbar\omega$ level density, we come to the
following relation:
\be\label{14}
\rho^{(0)}(E,J,1)=\rho(E,J,1)-\sum_{J'=|J-1|}^{J+1} \rho(E,J',0).
\ee

To calculate these partial level densities with the restricted values of excitation
numbers $N$ we need to slightly change the previous algorithm.
By construction, each configuration $\kappa$ in Eq. (\ref{1}) has a certain
excitation number $N_\kappa$ that can be defined as
\begin{equation}\label{15}
N_\kappa=\sum_{a=1}^n \nu_a- N_0,
\end{equation}
where the sum runs over all $n$ nucleons, $\nu_a$ is the excitation number of
the single-particle level occupied by the nucleon $a$, and, for our convenience,
we shift the result by the lowest value of the sum, $N_0$, so that $N_\kappa$
starts with zero: $N_\kappa=0, 1, 2, \cdots$. Knowing $N_{\kappa}$ we can
restrict the sum in Eq. (\ref{1}) by including only those configurations $\kappa$
which correspond to the desirable values of excitation numbers.

For each configuration $\kappa$ we need to calculate the $J$-fixed width and
centroid, $\sigma_{\alpha \kappa}$, $E_{\alpha \kappa}$, defined by Eqs.
(\ref{5}) and (\ref{6}). There is no problem with the centroid calculation,
$E_{\alpha \kappa} \sim \mbox{Tr}^{(\alpha \kappa)}[H]$, since it remains \
unchanged, we just need to select the configurations $\kappa$ of our interest.
The width calculation requires more attention. The problem is that the widths
depend on the trace of the second power of the Hamiltonian,
$\mbox{Tr}^{(\alpha \kappa)}[H^2]$, and if we want to restrict the basis we have
to take care of the intermediate states $-$ they also have to be restricted,
\be\label{16}
\mbox{Tr}^{(\alpha \kappa)}[H^2]=\sum_{\lambda \in \kappa} \sum_{\mu(N_{\mu} \in R)}
\langle\lambda|H|\mu\rangle\langle\mu|H|\lambda\rangle.
\ee
Here the sum over $\lambda$ includes all many-body states within the given
configuration $\kappa$ and with certain set of quantum numbers $\alpha$.
In contrast to this, the sum over $\mu$ includes all possible intermediate
states regardless to configurations and quantum numbers, but only restricted
by the desirable excitation numbers $N$. In Eq. (\ref{16}), $R$
represents the set of all such many-body states defined by the allowed $N$. 
In our previous algorithm, we
could remove the intermediate sum using the completeness relation,
$\sum_{\mu} \left|\mu \right> \left< \mu \right| = 1$, and express the width
in terms of the traces over single-particle excitations ($D-$structures, as in 
Eq. (11) of Ref. \cite{SH2010}). Now that is impossible, and we need to treat 
the situation differently.

One possible way to proceed here is still to follow the approach suggested in
Ref. \cite{jdenrc}, where the ``restricted" $D-$structures were introduced, see 
Eqs. (11,12) in this reference. 
%In the present work we apply a supposedly more effective way of dealing with the intermediate state restriction.
Here we propose an alternative solution, which may be easier to implement as an
efficient computer algorithm.
The single-particle part of the Hamiltonian (\ref{4}) does not create any problems.
Indeed the one-body operators of the $a^{\dagger}_{i}a_{i}$ type do not change
the excitation number $N$, so that all states $|\mu\rangle$ in Eq.
(\ref{16}) have the same $N_{\mu}$ as the states $|\lambda\rangle$ all of which
have the same excitation number defined by the configuration $\kappa$, $N_{\lambda}
=N_{\kappa}$. The two-body part of the Hamiltonian has the matrix elements
\begin{equation}
\langle\mu|V|\lambda\rangle=\,\frac{1}{4}\,\sum_{ijkl}V_{ijkl}
\langle\mu|a^{\dagger}_{i}a^{\dagger}_{j}a_{l}a_{k}|\lambda\rangle    \label{18}
\end{equation}
which indeed can mix the states with different excitation numbers.
%The idea is that instead of restricting the sum over $\mu$ in the eq.(\ref{16}) we replace the %Hamiltonian $H$ with a new operator $\tilde H$, which matrix elements satisfy the following %condition
%\be \label{17}
%\left< \nu \right| \tilde{H} \left|\mu \right> =\left\{
%\begin{array}{cll}
%\left< \nu \right| H \left|\mu \right> & ,& \mbox{if } \; \mu \in R \\
%0 & , & \mbox{if } \; \mu \not \in R
%\end{array}
%\right. .
%\ee
%Other words, restricted in such way Hamiltonian $\tilde H$ can lead to the transitions from %given $\left| \nu \right>$ state only into the restricted class of states $R$. Let us explain %how we construct this restricted $\tilde H$ Hamiltonian. First of all, the single-particle part %of $H$, see eq.(\ref{4}), does not need any restriction. Indeed, all operators of $a^\dag_i a_i$ %type do not change the excitation quantum number $N\hbar\omega$, and all the $\left| \mu %\right>$ states in eq.(\ref{17}) will have the same $N\hbar\omega$ as the $\left| \nu \right>$ %state. In opposite, the two-body part of $H$,
%\be \label{18}
%\sum_{i j k l} V_{i j k l} a^\dag_i a^\dag_j a_l a_k,
%\ee
%can and do mix the states of different excitation quantum numbers.
Each member of the sum in Eq. (\ref{18}) produces a certain change of the number of
excitation quanta, $\Delta N_{ijkl}$,
\begin{equation}
\Delta N_{ijkl}=\nu_{i}+\nu_{j}-\nu_{k}-\nu_{l},
\end{equation}
where $\nu_{i}$ is the excitation number of the single-particle level $i$.  It is
important that $\Delta N_{ijkl}$ reflects the internal property of the interaction
operator $a^\dag_i a^\dag_j a_l a_k$ and does not depend on the many-body states.
To deal with the sum over $\mu$ in Eq. (\ref{16}), we notice that $N_{\mu}=
N_{\lambda}+\Delta N_{ijkl}$, and therefore the restriction on $N_{\mu}$ can be
reformulated in terms of the equivalent restriction on the single-particle levels,
\begin{equation}
\sum_{\mu(N_{\mu}\in R)}\;\sum_{ijkl}\,\equiv \sum_{\mu}\left[\sum_{ijkl}\right]_{R_{\Delta}},
\end{equation}
where $R_{\Delta}$ means that the sum inside the parentheses is restricted by
$N_{\lambda}+\Delta N_{ijkl}\in R$.

In order to satisfy the condition of Eq. (17) we just need to keep those sets of
the single-particle levels $i, j, l,$ and $k$ in (\ref{18}) for which the excitation
quanta of the $|\lambda \rangle$ state plus the shift $\Delta N$, caused by the
corresponding operator $a^\dag_i a^\dag_j a_l a_k$, satisfy this restriction condition.
For example, if we want to take into account only $0\hbar\omega$ and $2\hbar\omega$
classes of states, and the $|\lambda \rangle$ states belong to the, let's say,
$0\hbar\omega$ class, then the allowed shifts of excitation quantum numbers will
be $\Delta N=0,2$. All other terms in Eq. (\ref{18}) must be ignored. Thus, we can
use the old expressions for the $D$-structures without changes (see Eq. (12) of Ref.
\cite{SH2010}), which is an essential advantage, but it is necessary to enforce the
appropriate restrictions on the sums over single-particle quantum numbers.

%=================
The final result can be written
as
%The traces in the restricted basis can be written in the following way
\begin{eqnarray} \label{19}
\nonumber
\mbox{Tr}^{(\alpha \kappa)}[H^2] = \\
\nonumber
=\sum_i \epsilon^2_i  D^{[i]}_{\alpha \kappa}
+ \sum_{i<j} \left[ 2 \epsilon_i \epsilon_j +  2 (\epsilon_i+\epsilon_j) V_{ijij}
\right] D^{[ij]}_{\alpha \kappa}\\
\nonumber
+\sum_{(i<j)\ne l} 2 \epsilon_l V_{ijij} D^{[ijl]}_{\alpha \kappa}
+\sum_{(i<j)\ne(q<l)} V_{ijij}V_{qlql} D^{[ijql]}_{\alpha \kappa}\\
\nonumber
%+\left(\sum_{i<j, \; q<l} V^2_{ijql} \left(D^{[ql]}_{\alpha \kappa}
%+ D^{[ijql]}_{\alpha \kappa}\right) \right.\\
+\left[ \sum_{i<j, \; q<l} V^2_{ijql} D^{[ql]}_{\alpha \kappa}
+\sum_{(i<j) \neq (q<l)} V^2_{ijql} D^{[ijql]}_{\alpha \kappa} \right.\\
\nonumber \left.
-\sum_{i, \; (q<l)\ne j} V^2_{ijql} D^{[jql]}_{\alpha \kappa} \right]_{R_1}\\
+\left[
\sum_{l, \;(i<j)\ne q} 2V_{liiq}V_{ljjq} \left( D^{[ijq]}_{\alpha \kappa}-
 D^{[ijql]}_{\alpha \kappa}  \right)
\right]_{R_2},
\end{eqnarray}
where $R_1$ means that the sums inside the associated square bracket are restricted by
$\left(N_\kappa+\nu_i+\nu_j-\nu_q-\nu_l\right) \in R$,
and $R_2$ means that the sums in corresponding square bracket are restricted by
$\left(N_\kappa+\nu_l-\nu_q\right) \in R$. The trace $D^{[i]}_{\alpha \kappa} = \mbox{Tr}^{(\alpha \kappa)}[a^\dag_i a_i]$
can be interpreted as a number of many-body states with fixed projection
$J_z$ (if we consider $J_z$-traces) and the single-particle
state $i$ occupied, which can be constructed for the configuration $\kappa$; the notations
for more complex traces are
$D^{[ij]}_{\alpha \kappa}= \mbox{Tr}^{(\alpha \kappa)}[a^\dag_i a^\dag_j a_j a_i]$,
$D^{[ijq]}_{\alpha \kappa} = \mbox{Tr}^{(\alpha \kappa)}[a^\dag_i a^\dag_j a^\dag_q a_q a_j a_i]$, and so on.
These $D$-structures were called propagation functions in Refs. \cite{SH2010,jaq79}.
The detailed procedure for calculating these functions can be found in Ref. \cite{SH2010}.

In some applications, the restriction of the level density to one class of excitations
described by the excitation number $N$, Eq. (10), might not be sufficient. For example,
one could be interested in considering the class of $(0+2)\hbar\omega$ excitations.
%The restrictions we used in our algorithm were made in the following way. In fact, we did not restrict the NLDs to the certain %$N\hbar\omega$. Instead of it,
In those cases we select the maximum value $N_{{\rm max}}$ of allowed excitations, and all the states, including intermediate ones, are restricted according to
\be\label{20}
N \leq N_{{\rm max}}.
\ee
Such a restriction allows us to fully take into account the interference between
the states of different $N\hbar\omega$ in Eq. (14). For example, if $N_{\rm max}=2$,
there are two possible classes of states, $0\hbar\omega$ and $2\hbar\omega$, contributing
to the width ($1\hbar\omega$ does not contribute because of opposite parity).
The cross terms in Eq. (14), which are proportional to
$ |\langle 2\hbar\omega | H |0\hbar\omega \rangle|^2$, are equally important along
with the diagonal contributions $ |\langle 2\hbar\omega | H |2\hbar\omega \rangle|^2$
and $|\langle 0\hbar\omega | H |0\hbar\omega \rangle|^2$.
%In this work we considered only cases with negative parity and $1\hbar\omega$ excitation class. There are no any cross terms %since $0\hbar\omega$ states have opposite parity, thus we could use eq.(\ref{12}) unchanged.
If one is interested in the density of unnatural parity states, one could consider only
the $1\hbar\omega$ excitations. In these cases there are no cross terms, since the
$0\hbar\omega$ states have opposite parity, thus one can directly use Eqs. (10)
and (12). Finally, one should mention that the present algorithm was integrated in our highly 
scalable moments code that was described in Ref. \cite{SH2010}.

\section{Results}

\begin{figure*}
\includegraphics[width=0.45\textwidth]{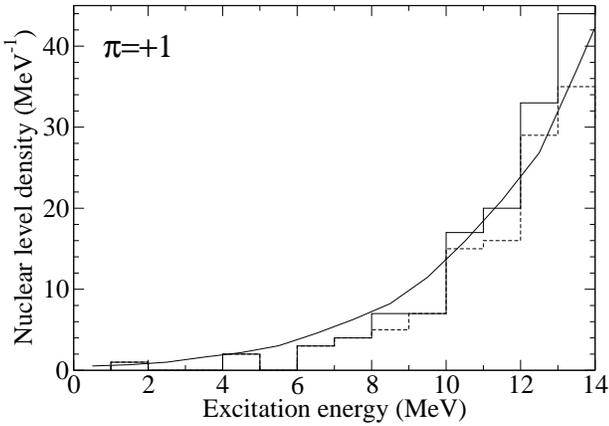}
\hfill
\includegraphics[width=0.45\textwidth]{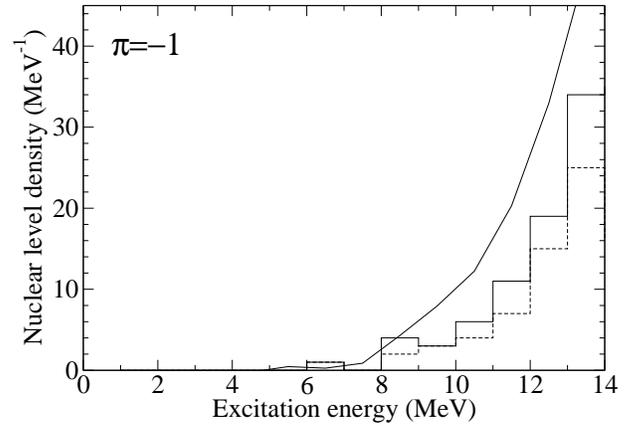}
\caption{${}^{28}$Si, all $J$, both parities. Comparison of nuclear level densities
from experiment (stair lines) and from the moments method calculation (straight line).}
\end{figure*}

As a first example we consider $^{20}$Ne in the $s-p-sd-pf$-shell model space.
For this space we use the Warburton-Brown (WBT) interaction \cite{wbt}. All calculations
were done for $1\hbar\omega$ subspace and for negative parity. Fig. 1 presents
the comparison of the exact shell model level densities (stair-dashed and stair-solid
lines) with the densities calculated using the moments methods (straight-dashed and 
straight-solid lines). The all dashed lines present the total densities including
spurious states. The solid lines present the pure non-spurious densities.
For the shell model calculations the spurious states were removed with the help of
the Lawson method \cite{lawson} that adds to the actual Hamiltonian a shifted
center-of-mass Hamiltonian, $H_{CM}$, multiplied by a constant $\beta$,
\be
H \rightarrow H'=H+\beta \left[\left(H_{CM}-\frac{3}{2}\hbar\omega\right)\frac{A}{\hbar \omega} \right].
\ee
The additional term pushes the all spurious states up and leaves the non-spurious density
at low-lying excitation energy. In our calculations we used $\beta=5$. However, as it
was mentioned in the introduction, the Lawson recipe can not filter spurious states for
the moments method. To get the straight-solid lines on Fig. 1 we use the recursive method 
introduced in Eqs. (10-12). Finally, the dotted lines present the spurious
densities itself calculated according to the second part of the right-hand side of
Eqs. (10,12). To calculate the level density with the moments method we need to know
the ground state energy and the cut-off parameter $\eta$. For $^{20}$Ne, the ground
state energy $E_{{\rm g.s.}}(^{20}\mbox{Ne})=-184.2$ MeV was calculated with the help
of the shell model, WBT interaction, $0\hbar\omega$ subspace, and $\eta=2.8$.

Figs. 2 and 3 present similar results for $^{22}$Na, $^{22}$Mg, in the $s-p-sd-pf$
model space. The calculations also were done with the WBT interaction, for
$1\hbar\omega$ subspace and negative parity. Stair lines refer to the shell model
calculations, while straight lines present the results of the moments method.
The only difference from Fig. 1 is the position of spurious states. The position
of the spurious contribution to the total level density calculated with the shell
model is naturally defined by the Lawson term and by parameter $\beta$. As we can
see from the Figures, $\beta=5$ is shifting all spurious states to the region of
excitation energies of order of 140 MeV. The spurious states calculated with
the moments method are mostly located near the ground state energy. To compare
the shape of the spurious part of the level density, we artificially shifted
the results of the moments method by energy $\beta A$ MeV, which is 110 MeV for
$A=22$ and $\beta=5$; after this, as seen from the Figures, the spurious states
calculated with the shell model and Lawson term almost completely coincide with
those calculated with the moments method and shifted afterwards. For all these
cases we chose $\eta=2.8$ and ground state energies $E_{{\rm g.s.}}=-201.2$ MeV for
$^{22}$Na and $E_{{\rm g.s.}}=-202.3$ MeV for $^{22}$Mg.

Figures 4 and 5 present bigger cases of $^{26}$Al and $^{28}$Si in the $s-p-sd-pf$
model space, for both positive and negative parities. The dimensions are very large, and
it is not practical to get the level densities with the shell model. Only the ground
state energies can be calculated for these cases in $0\hbar\omega$ subspace. With
the WBT interaction, we got $E_{{\rm g.s.}}=-250.3$ MeV for $^{26}$Al and
$E_{{\rm g.s.}}=-285.0$ MeV for $^{28}$Si. Figures 4 and 5 show comparisons of
the level density calculated with the moments method, that is presented by straight
lines, with experimental level densities presented by stair lines. There are
two stair-like lines on each figure: the solid stair lines present an ``optimistic"
approach, when all experimental levels with uncertain parity were counted; oppositely,
the dashed stair lines present a ``pessimistic" approach, when only experimental
levels with defined parity were counted.

It is needed to be mentioned that the real level densities must be greater than those
presented by the stair-like lines since it is possible that many levels were missed in 
experiment. In spite of the fact that the WBT interaction was not really tested in
such big model spaces (it was specifically designed for $A<20$), the agreement between
the calculations using the moments method and experimental data is quite remarkable.

\begin{figure*}
\includegraphics[width=0.45\textwidth]{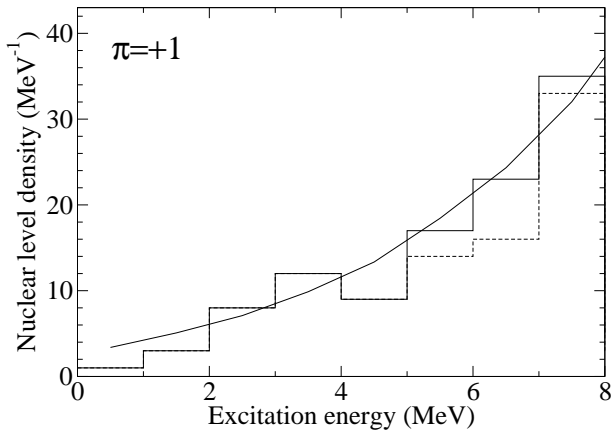}
\hfill
\includegraphics[width=0.45\textwidth]{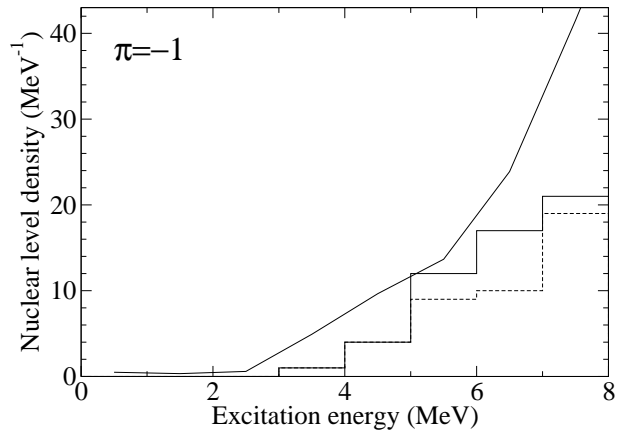}
\caption{${}^{26}$Al, all $J$, both parities. Comparison of nuclear level densities
from experiment (stair lines) and from the moments method calculation (straight line).}
\end{figure*}

%\begin{figure*}
%\includegraphics[width=0.45\textwidth]{mg241hw-11.eps}
%\hfill
%\includegraphics[width=0.45\textwidth]{mg241hw-12.eps}
%\caption{${}^{24}$Mg, parity=-1. Comparison of nuclear level densities
%between exact shell model (solid line) and moments method (dashed line). WBT interaction was used. }
%\end{figure*}

\section{Conclusions and Outlook}

In conclusion, we developed a new high-performance algorithm to calculate the configuration 
centroids and widths of the nuclear Hamiltonian in $N\hbar\omega$ subspaces built in a valence 
space. These first two moments can be used to calculate the SPNLD in the associated $N\hbar\omega$ 
subspaces, which can be further mixed according to a recently proposed algorithm \cite{dencom}
for extracting the non-spurious SPNLD. This strategy can be used to calculate the non-spurious 
shell-model level density for unnatural parity, a long-standing problem in nuclear structure.

We tested our techniques by calculating the negative parity level density for several even-even 
$sd$-nuclei, where the exact $1\hbar\omega$ shell model diagonalization can be done and compared 
with the results of the newly proposed algorithm. In all cases the SPNLD results of our moments 
method describe very well the results of the full shell model calculations. We also compared 
the results of our moments method with the available experimental data for $^{26}$Al and $^{28}$Si. 
For the states of positive parity, the effective interaction is well suited and our level
density compares successfully with the experimental data. For the negative parity states, there is 
no well-adjusted $1\hbar\omega$ effective interaction for the middle of the $sd$-shell.
Using the WBT interaction that was tested up to about mass 20 \cite{wbt} we obtain a reasonable 
description of the experimental data.

Certainly, there is a need of refined effective interactions for the unnatural parity states for 
the $sd$ nuclei, and our techniques could help improving their quality. The algorithm suggested 
in this article can be used to reliably predict for the first time the SPNLD for a large number 
of unstable nuclei relevant for the $rp$-process. The next steps could be directed to the more
complicated cases, where the $N\hbar\omega$ space is incomplete. This will allow us to reformulate
the approach for the realistic mean-field potentials when it will be possible to exclude the standard
reference to harmonic oscillator and open the broad field of problems related to the cross sections and 
reaction rates. In parallel, the deep question should be addressed of the influence of continuum 
effects \cite{celardo11} on the density of levels which are in fact resonance quasistationary states.   

%\begin{itemize}
%\item our old program was modified in order to perform NLD calculations within
%given $N\hbar\omega$ subspace.
%\item these partial NLDs allow us to exactly remove the center-of-mass spurious states.
%\item examples of NLD of, ${}^{20}$Ne, ${}^{22}$Na, ${}^{24}$Mg, in the $s-p-sd-pf$ model space are presented and compared with %Shell model calculations.
%\end{itemize}

%What else can (need to) be done:
%\begin{itemize}
%\item finish (or understand) the Shell model calculations for ${}^{24}$Mg.
%\item invent something for cases when $N\hbar\omega$ subspace is incomplete.
%\item applications (cross sections, reaction rates, ... ).
%\end{itemize}

\section{Acknowledgemnets}

R.S. and M.H. would like to acknowledge the DOE UNEDF grant No. DE-FC02-09ER41584 for support.
M.H. and V.Z. acknowledge support from the NSF grant No. PHY-0758099.

\end{document}